# International collaboration clusters in Africa

Jonathan Adams(1), Karen Gurney(1), Daniel Hook(2) and Loet Leydesdorff(3)

jonathanzadams@gmail.com

(1) *Evidence* Thomson Reuters, 103 Clarendon Road, LEEDS LS2 9DF, UK

(2) Symplectic, 10 Crinan St, LONDON N1 9XW, UK

(3) Amsterdam School of Communication Research (ASCoR), University of Amsterdam, Kloveniersburgwal 48, 1012 CX Amsterdam, The Netherlands

## Abstract

Recent discussion about the increase in international research collaboration suggests a comprehensive global network centred around a group of core countries and driven by generic socio-economic factors where the global system influences all national and institutional outcomes. In counterpoint, we demonstrate that the collaboration pattern for countries in Africa is far from universal. Instead, it exhibits layers of internal clusters and external links that are explained not by monotypic global influences but by regional geography and, perhaps even more strongly, by history, culture and language. Analysis of these bottom-up, subjective, human factors is required in order to provide the fuller explanation useful for policy and management purposes.

## Introduction

Georghiou (1998) drew attention to the phenomenon of increasing co-authorship in research publications and, noting previous studies including Frame & Carpenter (1979), associated this particularly with major global research facilities and cooperative programmes. He identified a number of barriers to the potential growth of 'bottom up' cooperation but concluded that formal 'top down' enabling arrangements were emerging through, for example, the European Framework Programmes (FPs). Persson et al (2004) noted deepening scientific collaboration and increasing citation impact in all science areas. In particular, authorship was increasing exponentially while the number of collaborative papers was increasing linearly so connectedness was intensifying. Greene (2007) confirmed this trend and King (2012) described the growing frequency of massively multi-authored papers.

Wagner & Leydesdorff (2005) argued that patterns in international collaboration in science can be considered as network effects and that only the European FPs noted by Georghiou (op. cit.) mediated relationships at that level. Their global network shares features with other complex adaptive systems in which order emerges from interactions between many agents pursuing self-interested strategies. Adams, Gurney & Marshall (2007) pointed to the intense levels of interactions between leading research economies. Leydesdorff & Wagner (2008) suggested that the global network reinforced a core group of (fourteen) cooperative countries with strong national systems. They argued that peripheral countries could be disadvantaged by increased strength at the core.

Wagner (2008) argues, from complex systems theory, that the self-organizing global system influences all lower systems (Wagner et al, in prep). Here, we accept the meta-pattern but contest the network as a sufficient explanatory model and concur with Georghiou, that there are other agents such as major facilities (e.g. CERN – see King, 2012) and cooperative programmes (e.g. WHO,

FAO, climate change) which have been important. In addition, we argue that the effects of history, culture and language continue to have a profound human influence on collaboration patterns, mediated through personal preference rather than strategic logic (Adams, 2012).

In this paper we illustrate these effects through an analysis of collaboration patterns in Africa (see also Adams, King & Hook, 2010). Africa contains more than 50 nations, hundreds of languages, and a welter of ethnic and cultural diversity. OECD's African Economic Outlook (OECD et al., 2011) sets out in stark detail the challenge for the research base in Africa and the extent to which current global economic problems may make this worse and further compromise the value of the commitment made in by developed nations 2005 to double official development assistance to Africa by 2010. More than half the African nations are off-track or regressing on objectives to achieve universal primary education by 2015. Internet penetration is good only in North Africa, constraining communication and access to knowledge. It needs international research partners.

Is the uniform, generic pattern perceived elsewhere also found in Africa, or does the continent exhibit more subtle influences in its patterns of research collaboration? And, picking up the visualisation methods compared by Leydesdorff et al (in press), how can we best represent what we see?

## Methods

We focussed on research publications with one or more addresses for a country within Africa as defined by the UN. We sampled data for the period 2000-2012 (data to current indexing, not year-end). We also collated data on GDP for each country for which publication data were available.

Volume and subject analyses used Thomson Reuters *National Science Indicators*. Collaboration analyses were carried out using *Research Performance Profiles* data in *InCites™*, a web-based platform for research evaluation from Thomson Reuters. Database years were used to delineate years, and only article, note and review document types were considered.

We counted all collaborations between countries represented by co-authorship on the publications we collated. The counts are by paper not by number of researchers. For example, a paper co-authored by two researchers from Ghana, three from Nigeria and one from Kenya counts as a single paper in each country's total and as one link between each pair of countries.

Analysis was extended using *Wolfram Mathematica® 7* to create maps and collaboration diagrams. We also had access to the data collected about 2011-publications by Leydesdorff et al. (in press), and extracted the subset of African countries.

## Results

Total research output for Africa increased from 13,271 publications indexed on Thomson Reuters Web of Science in 2000 to over 35,000 publications in 2012 (34,528 catalogued at Dec 15, 2012). For reference purposes, the total output of Africa is about the same as that of the Netherlands.

The percentage of Africa's publications that were substantive research papers (that is to say, articles or reviews) declined from 88% to 82.6%, which reflects an increasing number of proceedings papers and other contributions authored within Africa. (Figure 1)

The number of articles and reviews that have been authored wholly within Africa (i.e. that have no collaborative co-author from outside the region) has doubled since 2000 from 6,319 to 12,089. This is a decline as a percentage of total research paper output from 54% to 42%. However, the relative collaborative output of G8 countries rose much faster over the period: collaboration is increasing everywhere. Thus, in fact, the autonomous research output of Africa clearly grew in the last decade and Africa is becoming increasingly self-reliant in this regard.

A breakdown of the figures demonstrates the extent to which each region—and African science as a whole—is dominated by three nations: Egypt in the north, Nigeria in the middle, and South Africa in the south. In this millennium, since 2000, Egypt produced nearly 58,000 publications which was more than twice the total for Tunisia, its next-place and regional neighbour. In west-central Africa, Nigeria's total for the same period was over 20,000, compared to roughly 12,800 for Kenya which is the leading research economy in the east of the continent. South Africa's dominance, as might be expected, is even more pronounced: over 95,000 publications since 2000, compared to the southern region's next-most-prolific nation, Tanzania, which fielded just over 6,300. (Figure 2a, Table 1)

What happens when we break the publication data down by field of research? In our recent Global Research Report on Africa we showed a discernible pattern in Africa's relatively high representation—as a share of world publications—in fields that are relevant to natural resources. The highest percentage of any field, for example, is South Africa's 1.55% share of Plant & Animal Science. Not far behind is the same country's 1.29% share of Environment/Ecology. A review of the more detailed analyses in Thomson Reuters *Essential Science Indicators* shows that many of South Africa's most highly cited papers in this field pertain to climate change and its effects on plant propagation. Following this theme, South Africa's 1.13% share of Geosciences is in keeping with the region's mineral richness.

In short, Africa is a continent abundant in natural resources. How much does Africa itself benefit from those resources? Absolute volume of published papers is one indicator of research activity and—indirectly—of research capacity. It will therefore be obvious that the output of a country reflects how much money is going in to its research system, and that is likely to be partly dependent on its general economy. Bigger countries with a larger economy should be producing more papers, if they invest at the same level as smaller countries. However, land area, population density and resources vary a great deal. We have compared publications with Gross Domestic Product (GDP) for each country, reasoning that proportionate investment in the knowledge economy is a good index of a government's commitment to maximize the longer term benefit of resource development and exploitation for the general wealth of its people.

The leading countries by output are South Africa, Egypt, Nigeria, Tunisia, Algeria and Kenya (Figure 2a). Four of these (South Africa, Egypt, Nigeria and Algeria) are also leading countries in terms of GDP while Kenya and Tunisia fall in a lower GDP tier. Indexing output against GDP (Figure 2b) provides further interpretation. Zimbabwe is highlighted as relatively the most productive country in terms of publications per unit GDP but this is anomalous because it retains its legacy research base despite a collapsing economy and very low current GDP. The real leaders are Tunisia and Malawi with very different economic bases but strong relative productivity in both cases. South Africa, Kenya and Egypt all have significant relative productivity, as do a number of other countries in East Africa (Ethiopia, Uganda, Tanzania) and West Africa. (Cameroon, Ghana).

It is clear, however, that despite Nigeria's high volume output it is not producing as much research as would be expected given the size of its economy. The value of its resources is not yet being felt in its knowledge base. In fact, the same research productivity gap between potential and actual investment applies to several other countries. This is an area where Africa is not yet benefitting from the best use of its own natural resources.

Africa's research can be boosted by collaborative international partnerships. The countries collaborating most frequently with partners in Africa are the USA (39,292 papers between 2000 and 2012), France (31,421), the UK (25,753), Germany (13,879) and Canada (7,604). This looks like a roll-call of 'the usual suspects' among major research producers. It is therefore worth noting that Saudi Arabia collaborated on 6,285 papers, albeit almost entirely with countries in North Africa of which Egypt (4,939 joint publications) was the pivotal link. Ethiopia's research base is distinctive in being substantial, growing and yet almost entirely domestic. The most substantial links between countries in Africa and the USA, UK, France, Saudi Arabia are summarised in Table 1.

The research axis between Egypt, Saudi Arabia and the USA is an instructive example of new and changing collaboration patterns. The numbers of papers co-authored between Egypt and the USA has grown but has remained around 10% of Egyptian output since 1995. The numbers of papers co-authored between Egypt and Saudi Arabia has been much smaller historically but reached 100 (4% of Egypt's output) in 2002 and exceeded 1000 (15%) in 2011. This is regionally, not globally, driven: only a small fraction of these papers also have the USA as a co-author. (Table 2)

France also has a niche relationship with Africa. It is unusual in studies of international collaboration to find it high in any ranking, and here to be 2nd behind the USA, ahead of the UK and with much more than twice the collaboration links of Germany. Among the 31,421 total co-authorships by partner then we find that Tunisia (Table 1: 7,400 23.6%) leads with Algeria (19%), Morocco (18.3%) and then Cameroon (5.5%). France is focussed on a small set of countries just across the Mediterranean in North Africa.

The USA and the UK, by contrast, collaborate diversely with South Africa, Kenya, Egypt, Nigeria and others (Table 1). The UK has much greater collaboration with specific countries, such as the Gambia and Malawi, than any other partner. Clearly, no collaboration pattern in Africa is general or uniform.

For each of six key research economies in Africa we have analyzed collaborative research links by collating co-authorships with other countries and analyzing collaboration with the USA and the UK as the most frequent partner for most countries, and three other frequent partners. (Figure 3)

How can we best visualise research co-publication within Africa? First, to create a simplified picture useful as an indication of major links for policy purposes, we used a threshold to clarify where relatively strong and persistent collaborations occurred. The threshold was set at a minimum of five papers per year, or 25 papers in total over the recent five-year period. This meant that some countries did not appear at all in the analysis because they had too low a level of recent collaboration. We then used a grouping algorithm to associate the countries around the rim of the wheel until groups with strongest cross-links were placed close together. (Figure 4)

Second, we created a more complete but necessarily more complex picture of the entire Africa network 2011 using VOSViewer. (Figure 5)

## Discussion

This analysis presents a complex picture of diverse research collaboration links, internationally (Table 1, Figure 3) and within Africa (Figure 4, Figure 5). It is difficult to argue that these outcomes are a response to a common global network phenomenon rather than local, cultural and historical factors that play into research opportunities and create the highly individualistic and specific African outcome. We do not disagree with the concept that international research collaboration is a common phenomenon but we do believe in the need to determine the bottom-up regional and local factors that properly explain complex outcomes departing from a notional top-down global template. Only by understanding this detail will research performance analysis engage with the theory and practice of research policy and management.

The research output of Africa is growing although remains small compared to established economies (Figure 1). Africa has enormous natural resources but, while there is a broad relationship with investment as GDP (Figure 2), some richer countries have yet to commit to substantial investment in their knowledge economy. It is therefore to be anticipated that further research development will continue to benefit from extensive external support and collaboration.

External collaborative links vary significantly by country. France is the second most frequent collaborator with Africa, after the USA, with concentrated links to North-West Africa and to central West Africa. It is interesting to note that, after normalization for size, Leydesdorff & Wagner (2008: 321) found France as highest on betweenness centrality because of its intermediating function with the EU networks. The UK is the most frequent collaborator with other African countries, such as Malawi and Gambia (Table 1). These links are not driven by global phenomena but by local historical and cultural factors and by targeted international cooperative health and food programmes. Many links are mediated through cooperative health and agricultural programs. Gambia is the site for long-term research into tropical diseases for the UK's Medical Research Council (Adams, Gurney & Pendlebury, 2012) which also works in Uganda. The Wellcome Foundation has similar, major research investments in Kenya and Malawi. A significant intellectual benefit is thus secured outside Africa.

Another exceptionally strong link is that between Egypt and Saudi Arabia, which is not mediated by a third party such as the USA (Table 2) but through their axis in supporting regional growth in research capacity in the Arabian Middle East. (Figure 3; Adams et al, 2011)

How can we create a picture of Africa's research network that would be helpful for policy engagement? If we apply a threshold on the strength of interaction we find no single network within Africa. Interface with African countries requires awareness that collaboration is driven partly by geography but also by shared culture and—very strongly—by language. (Figure 4)

(1) There is a marked interaction between the countries in North Africa which share both language and culture and are also relatively prolific. Thus, this network is probably the strongest group overall since it links countries which are individually research active across multiple fields. The group does little research with the rest of Africa, however, other than through an Egypt-South Africa link.

(2) A West Africa group (Benin-Togo) pivots around Cameroon, a relatively research productive country. The common factor within this group is almost certainly their common use of French as the cross-national business language.

(3) Language also gives us the clue to the large group which includes Kenya and geographical neighbours in East Africa but also includes Nigeria, Ghana and Gambia. Those countries appear to have English as a common language or have had a strong Anglophone influence.

(4) The Southern African Development Community (SADC) does not emerge as a research network since it is split between that group linked to Kenya and Nigeria and a second group most closely linked to South Africa, but which also includes Sudan and Gabon. The overall collaboration network, to the extent that one exists at all, is dependent on a small number of key players linking these regional and cultural groupings.

The simplified collaboration cartwheel of Figure 4 is useful for managers and planners. It is expanded and developed in Figure 5 into a complementary visualisation where completeness adds complexity requiring additional interpretation. It is therefore of greater value for the analyst. The map highlights the pivotal role of South Africa: the research hub in every sense. The map shows that Egypt is not embedded in the separated cluster of North Africa but is an outlier due to its wider attachments. There is a strong East Africa group, as in Figure 4, but there is also the development of two distinct groups in West Africa.

In Figure 3, there is a striking difference between the countries pulled out in North Africa (Figure 5) and those in other regions. Globally, the most frequent collaborative partner is the USA. Often this is a consequence of researchers who have studied in the USA maintaining links with those research groups when they return home. The UK and Germany are the other common partners to the countries featured here and France has a major role. This is the influence of the global network (pace Wagner *op. cit.*): between them the USA, UK, Germany and France have authors on half the world's research papers every year.

Nigeria sits at a research crossroads between East, West and South Africa. Despite its disappointing level of research investment, it has an important connecting role. Not only is it a part of the Anglophone collaborative network but it also has significant—albeit weaker—connections with its West African neighbours, and it connects strongly to South Africa. South Africa is a similarly strong node with a spread of links into other groups. These two, with Kenya, create strong cross-continent links and are key nodes into global networks.

China and Brazil's rapidly expanding research bases collaborate only weak with Africa. Nigeria's global reach is marked by some collaboration with China. It is theoretically well-positioned to extend its links westwards and partner with the emerging Brazilian research base. It could thus serve as a key doorway into both the West African and the Anglophone African research base for some of the exciting research which is now appearing in Asia and Latin America. But it has yet to realise this opportunity.

Table 1 Research output and collaboration (all publications on *Web of Science*, 2000-2012) between the USA, UK, France, Saudi Arabia and their most frequent partners in Africa

| | | USA | France | UK | Saudi Arabia |
|---|---|---|---|---|---|
| Africa - Total | 296,351 | 39,292 | 31,421 | 25,753 | 6,285 |
| South Africa | 95,309 | 14,264 | 3,801 | 10,131 | |
| Egypt | 57,741 | 5,900 | 1,019 | 2,409 | 4,939 |
| Kenya | 12,769 | 4,260 | 460 | 2,791 | 18 |
| Uganda | 6,317 | 2,318 | 231 | 1,402 | 11 |
| Nigeria | 21,909 | 1,945 | 243 | 1,426 | 54 |
| Tanzania | 6,299 | 1,693 | 171 | 1,625 | 8 |
| Ghana | 4,945 | 1,159 | 156 | 1,003 | 14 |
| Malawi | 2,909 | 990 | 124 | 1,087 | |
| Morocco | 17,518 | 956 | 5,738 | 559 | 186 |
| Ethiopia | 5,579 | 933 | 218 | 576 | 26 |
| Cameroon | 5,915 | 832 | 1,730 | 548 | |
| Tunisia | 24,724 | 755 | 7,400 | 421 | 326 |
| Senegal | 3,634 | 573 | 1,622 | 275 | 3 |
| Algeria | 14,846 | 412 | 5,961 | 292 | 367 |
| Gambia | 1,294 | 297 | 86 | 857 | |
| Gabon | 1,188 | 241 | 504 | 205 | |

Table 2 Growth of Egypt's research output and its collaboration with the USA and with Saudi Arabia over thirty years from 2000-2012 (part year). Egypt has increased collaboration with Saudi Arabia and little of this is driven by its prior links with the USA.

| Year | Egypt total | Egypt + USA | USA as % Egypt | Triple co-authors | Saudi as % Egypt | Egypt + Saudi |
|---|---|---|---|---|---|---|
| 2000 | 2,577 | 286 | 11 | 2 | 4 | 95 |
| 2001 | 2,707 | 227 | 8 | 3 | 3 | 94 |
| 2002 | 2,894 | 295 | 10 | | 4 | 115 |
| 2003 | 3,238 | 312 | 10 | 7 | 6 | 181 |
| 2004 | 3,212 | 318 | 10 | 4 | 5 | 169 |
| 2005 | 3,338 | 326 | 10 | 3 | 5 | 164 |
| 2006 | 3,847 | 358 | 9 | 6 | 5 | 190 |
| 2007 | 4,280 | 424 | 10 | 8 | 5 | 199 |
| 2008 | 4,710 | 439 | 9 | 15 | 6 | 261 |
| 2009 | 5,725 | 597 | 10 | 20 | 7 | 416 |
| 2010 | 6,281 | 708 | 11 | 33 | 10 | 614 |
| 2011 | 7,416 | 823 | 11 | 55 | 15 | 1,093 |
| 2012 (part) | 4,386 | 428 | 10 | 47 | 19 | 832 |

Figure 1  Africa’s output of publications indexed on Thomson Reuters *Web of Science* databases between 2000 and 2012

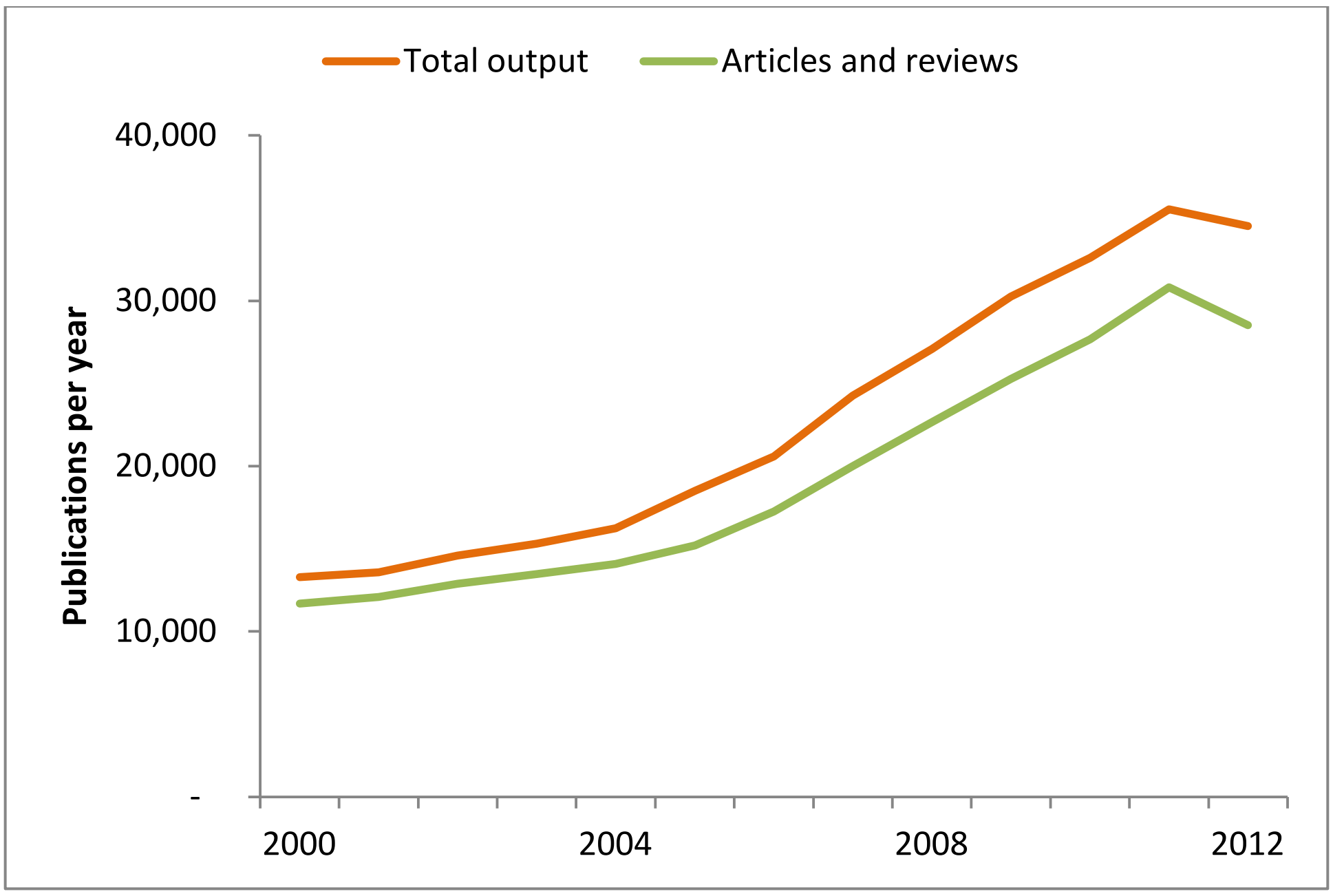

Figure 2.  Output per country in 2008 as total volume (Figure 2a) and as volume/GDP (Figure 2b). South Africa is absolutely the most productive country.  Zimbabwe appears to be relatively productive but this is an anomaly due to very low recent GDP and a strong historical base.  Tunisia is relatively the more productive on current performance.

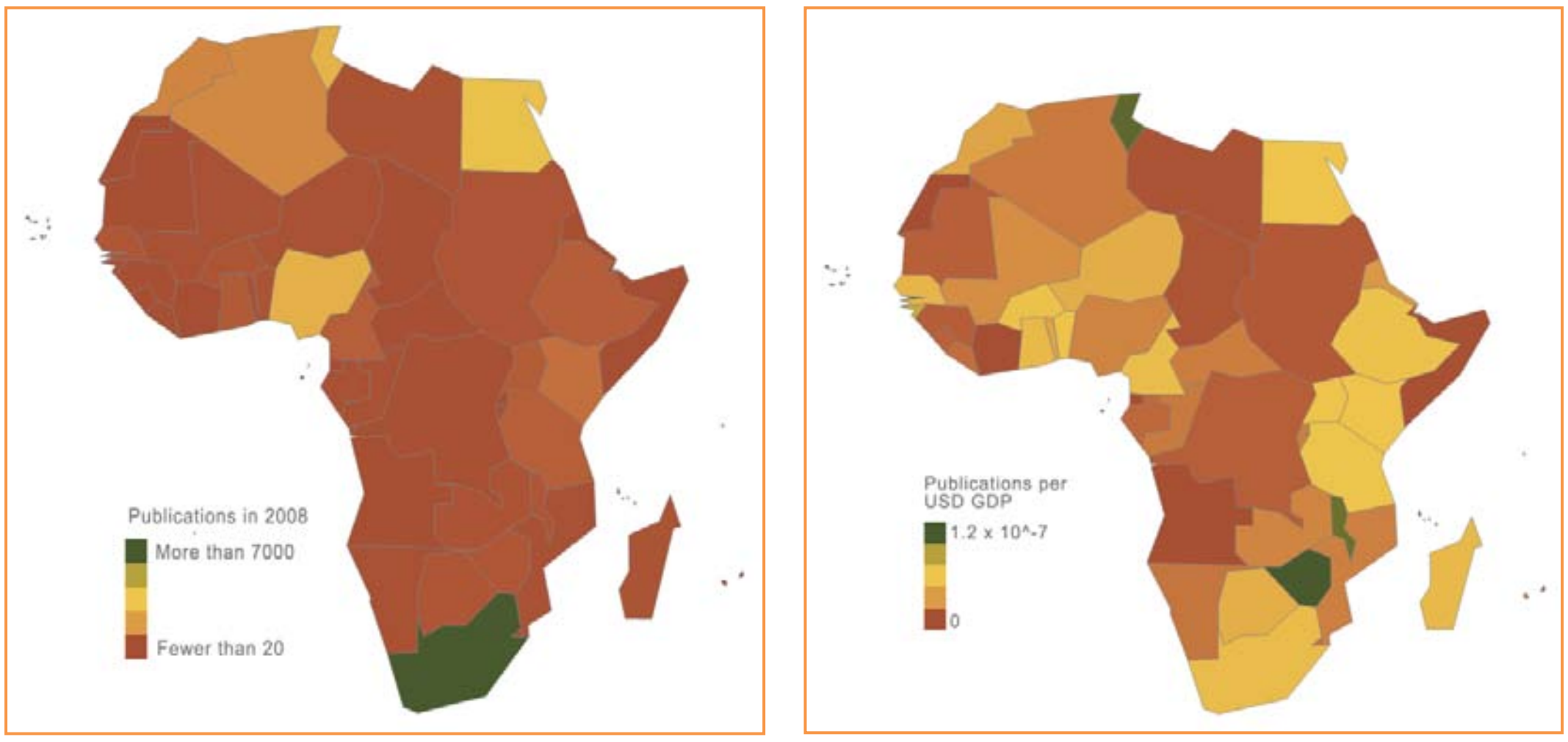

Figure 3  Most frequent intercontinental research collaborations for six key African research economies

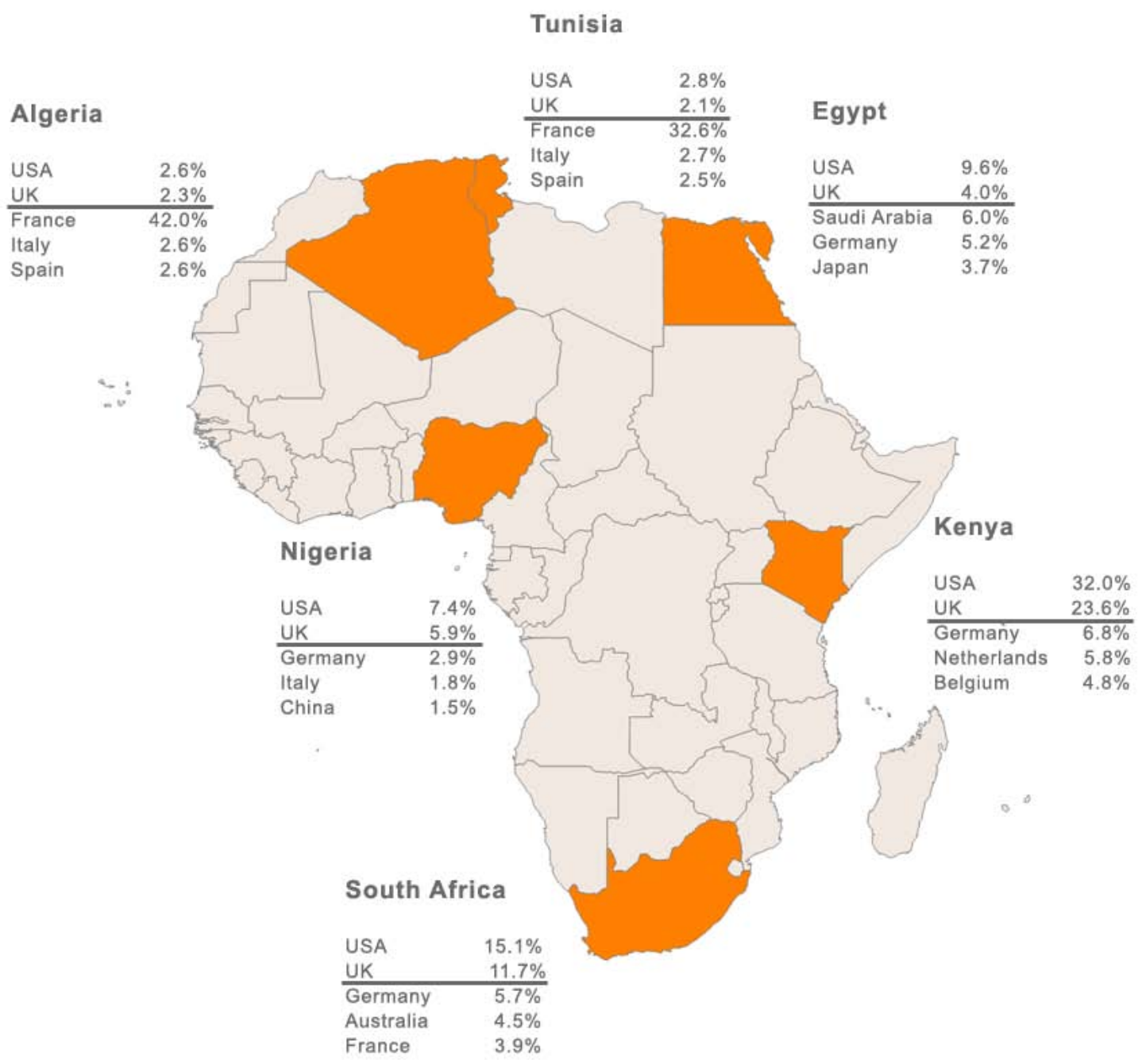

Figure 4: Collaboration between countries within Africa. This dependency graph uses *Wolfram Mathematica® 7 to* provide a new visual interpretation of collaboration, by paper not by number of researchers, and reveals clusters of countries with strong and persistent partnerships. Links displayed between each country meet a threshold of five publications per year for a continuous period of five years.

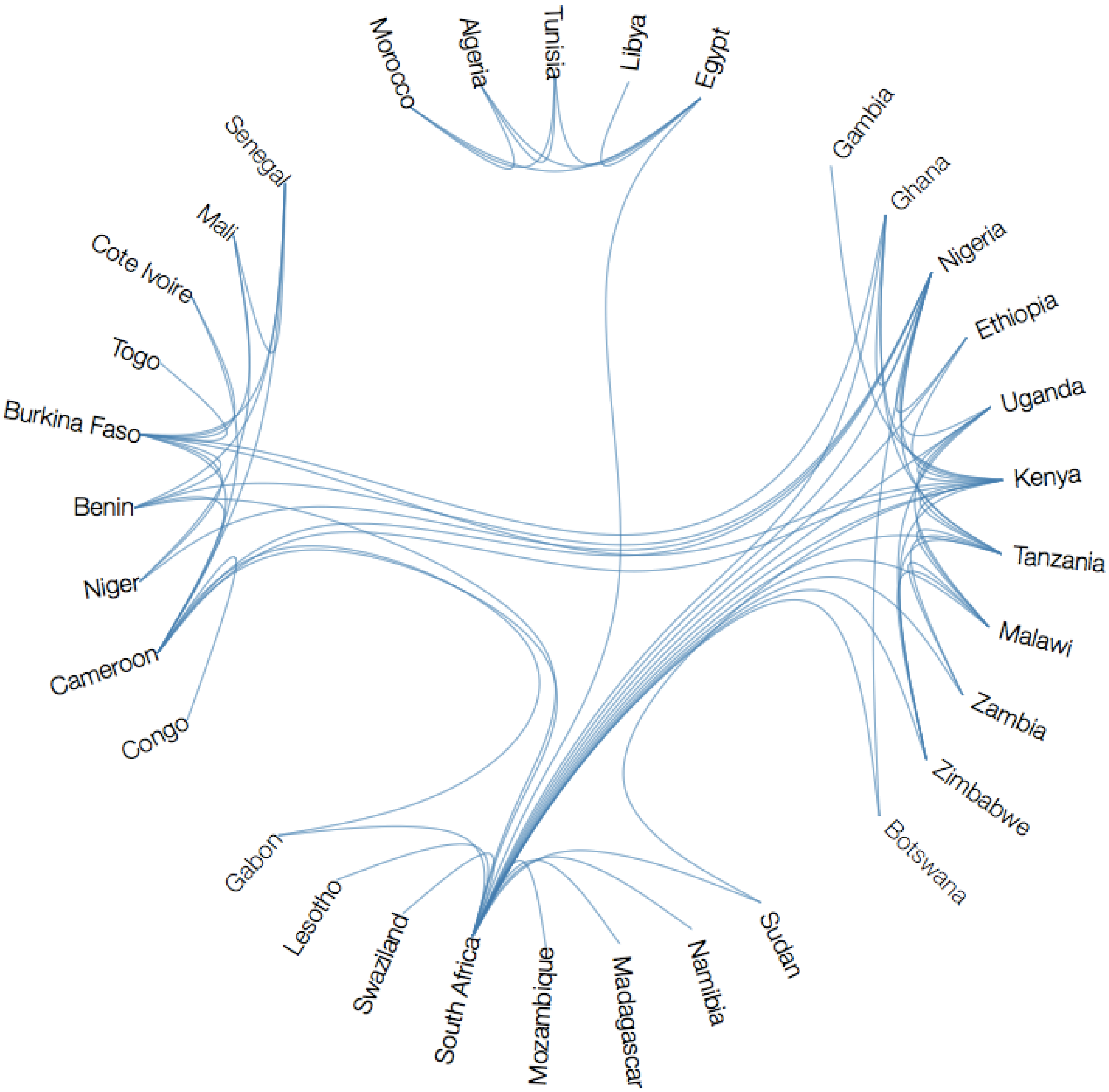

Figure 5: Coauthorship relations between 46 countries within Africa in 2011. VOSViewer was used for the clustering and mapping. This map highlights the pivotal role of South Africa, shows the separated cluster of North Africa with Egypt as an outlier due to its wider attachments, and recognizes not only the East Africa group but also the development of two distinct groups in West Africa. The map can be web-started at [http://www.vosviewer.com/vosviewer.php?map=http://www.leydesdorff.net/intcoll/afr_map.txt&network=http://www.leydesdorff.net/intcoll/afr_netw.txt&label_size=1.25&n_lines=1000](http://www.vosviewer.com/vosviewer.php?map=http://www.leydesdorff.net/intcoll/afr_map.txt&network=http://www.leydesdorff.net/intcoll/afr_netw.txt&label_size=1.25&n_lines=1000) for further exploration.

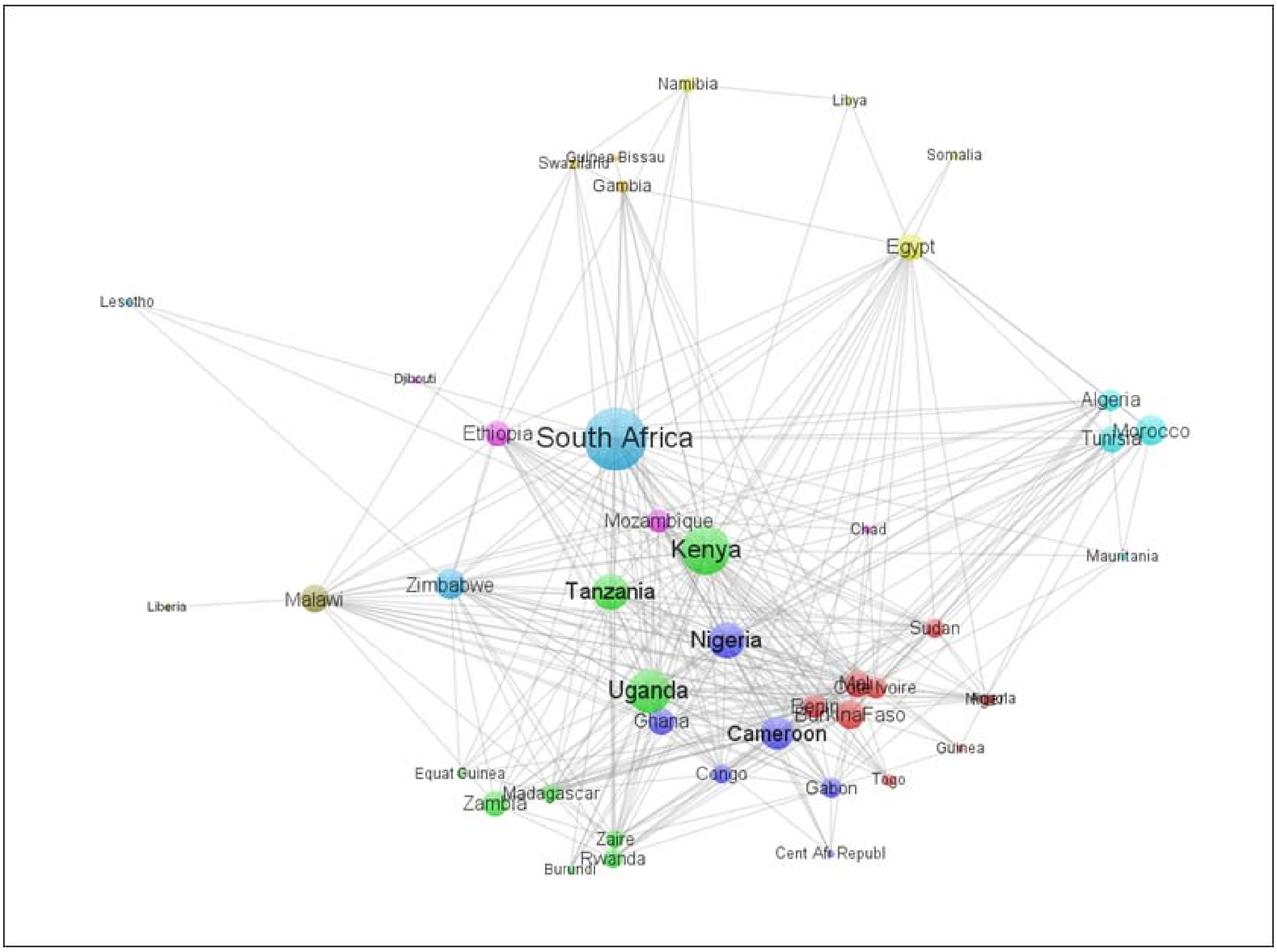